\documentstyle[aps,prl,preprint,tighten,epsf,floats,eqsecnum]{revtex}

\begin{document}
\title{Distribution of the quantum mechanical time-delay matrix for a
chaotic cavity}
\draft

\author{P. W. Brouwer}

\address{Lyman Laboratory of Physics, Harvard University,
Cambridge, MA 02138, USA}

\author{K. M. Frahm}

\address{Laboratoire de Physique Quantique, UMR 5626 du CNRS,
Universit\'e Paul Sabatier, 31062 Toulouse Cedex 4, France}

\author{C. W. J. Beenakker}

\address{Instituut-Lorentz, Leiden University, P.O. Box 9506, 2300 RA
Leiden, The Netherlands}

\date{\today}

\maketitle

\begin{abstract}
We calculate the joint probability distribution of the Wigner-Smith
time-delay matrix $Q=-i\hbar\,S^{-1} \partial S/\partial \varepsilon$
and the scattering matrix $S$ for scattering from a chaotic cavity
with ideal point contacts. Hereto we prove a
conjecture by Wigner about the unitary invariance property of the
distribution functional $P[S(\varepsilon)]$
of energy dependent scattering matrices $S(\varepsilon)$.
The distribution of the inverse of the eigenvalues $\tau_1,\ldots,\tau_N$
of $Q$ is found to be the Laguerre ensemble from random-matrix theory.
The eigenvalue density $\rho(\tau)$ is computed
using the method of orthogonal polynomials. This general theory
has applications to the thermopower, magnetoconductance, and capacitance
of a quantum dot.
\end{abstract}


\section{Introduction}
\label{sec1}

The time delay in a quantum mechanical scattering problem is
related to the energy derivative of the scattering matrix
$S(\varepsilon)$. Although the notion of such a relationship
goes back to the historical works of Eisenbud \cite{eisenbud}
and Wigner \cite{wigner1}, the
first to propose a matrix equation linking time delay and
scattering matrix was Smith \cite{Smith}. He introduced the hermitian
matrix $Q = -i \hbar S^{-1} \partial S/\partial \varepsilon$, now
known as the Wigner-Smith time-delay matrix, and interpreted
its diagonal elements as the time delay of a wave packet
incident upon the scatterer in one of $N$ scattering channels.
The eigenvalues $\tau_1,\ldots,\tau_N$ of $Q$ are referred to
as the ``proper delay times''.

In recent years, the issue of time delay and its connection to
the scattering matrix has received considerable attention
in the context of chaotic scattering
\cite{FSrev,Brev,Lyu,Bau,LW,Har,Eck,Izr,Leh,FS1,SZZ,GMB,BB,BFB,MJP,CT}.
Examples are the scattering of electromagnetic or sound waves from
chaotic ``billiards'' or the transport of electrons through so-called
``chaotic quantum dots'' \cite{qdreview}. In both cases the
scattering matrix $S$ and the time-delay matrix $Q$ have a
sensitive and complicated dependence on microscopic
parameters, such as the energy $\varepsilon$, the
shape of the cavity, or (in the electronic case) the magnetic
field. As a result, a statistical approach is justified,
in which one addresses the
statistical distribution of $S$ and $Q$ for an ensemble of
chaotic cavities,  rather
than the precise value of these matrices for a specific
cavity.

For the scattering matrix $S$, there is considerable theoretical
\cite{LW,BlumelSmilansky1,BlumelSmilansky1b,BlumelSmilansky2,Brouwer} and
experimental \cite{Smilansky,microwave1,microwave2,microwave3}
evidence that its distribution is uniform in the group of
$N \times N$ unitary matrices, restricted only by fundamental symmetries,
if the cavity is coupled to the outside world via ballistic point contacts.
This universal distribution is Dyson's
circular ensemble of random-matrix theory \cite{Dyson-4,Mehta}. The
fundamental problem is to calculate the distribution of the time-delay
matrix $Q$ for this case of ideal coupling. Known results include
moments of $\mbox{tr}\, Q$ \cite{FSrev,Leh,FS1} and the entire
distribution for the case $N=1$ \cite{FS1,GMB} of a scalar scattering
matrix.

In a recent paper \cite{BFB} we succeeded in computing the entire
distribution for any dimensionality $N$ of the scattering matrix. We
found that the eigenvalues $\tau_1,\tau_2\ldots,\tau_N$ of $Q$ are
statistically independent of the scattering matrix $S$, with a distribution
that is most conveniently expressed in terms of the inverse
delay times $\gamma_n = 1/\tau_n$,
\begin{equation}
  P(\gamma_1,\ldots,\gamma_N) \propto
  \prod_{i<j} |\gamma_{i} - \gamma_{j}|^{\beta}
  \prod_k \gamma_{k}^{\beta N/2}
  e^{-{\beta \tau_H \gamma_k/2}}. \label{eq:Lag}
\end{equation}
Here $\beta=1,2,4$ depending on the presence or absence
of time-reversal and spin-rotation symmetry, $\tau_H = 2 \pi \hbar/\Delta$
is the Heisenberg time, and $\Delta$ is the mean level spacing of the
closed cavity. The function $P$ is zero if any one of the $\gamma_n$'s
is negative. This distribution is known in random-matrix theory as the
(generalized) Laguerre ensemble \cite{Mehta}. The eigenvectors of $Q$
are not independent of $S$, except in the case $\beta=2$ of broken
time-reversal symmetry. For any $\beta$ the correlations are transformed
away if $Q$ is replaced by the symmetrized matrix
\begin{equation}
  Q_\varepsilon = -i \hbar\, S^{-1/2} {\partial S \over \partial
  \varepsilon} S^{-1/2},
  \label{eq:QE}
\end{equation}
which has the same eigenvalues as $Q$. The matrix of eigenvectors $U$ which
diagonalizes
$Q_\varepsilon = U \mbox{diag($\tau_1,\ldots,\tau_N$)} U^{\dagger}$ is
independent of $S$ and the $\tau_n$'s, and uniformly distributed in the
orthogonal, unitary, or symplectic group (for $\beta=1$, $2$, or $4$,
respectively).

In this paper we present a detailed and self-contained derivation of
these results, focusing on those parts of the derivation that have
not or only briefly been discussed in Ref.\ \onlinecite{BFB}. We
feel that a detailed presentation of this derivation is important,
because it relies on an old conjecture by Wigner \cite{wigner2} that
had not been proven before in the literature. (An unpublished proof
is given in Ref.\ \onlinecite{unpublished}.) The conjecture concerns
the invariance of the distribution functional $P[S(\varepsilon)]$
of the ensemble of energy dependent scattering matrices under unitary
transformations $S(\varepsilon)\rightarrow VS(\varepsilon)V'$ (with
$V,V'$ two energy-independent unitary matrices). Our proof of Wigner's
conjecture is based on the Hamiltonian approach to chaotic scattering
\cite{vwz}, which connects the scattering matrix $S(\varepsilon)$ of
the open cavity with leads to the Hamiltonian ${\cal H}$ of the closed
cavity without leads. It describes the so-called ``zero-dimensional'' or
``ergodic'' limit in which the time needed for ergodic exploration of the
(open) cavity is much smaller than the typical escape time.  In Ref.\
\onlinecite{Brouwer} one of us used the Hamiltonian approach to derive
the unitary invariance of the distribution {\em function\/} $P(S)$
of the scattering matrix at a fixed energy. This unitary invariance is
known as Dyson's circular ensemble \cite{Dyson-4}. Wigner's conjecture
is the much more far-reaching statement of the unitary invariance of
the distribution {\em functional\/} $P[S(\varepsilon)]$.

The outline of this paper is as follows. In Sec.\ II we formulate Wigner's
conjecture in its most general form for arbitrary dimensionality $N$
of the scattering matrix. (Wigner only considered the scalar case
$N=1$.) The proof follows in Sec.\ III. In Sec.\ IV we show how the
distribution of the matrices $S$ and $Q$ follows from this, now proven,
conjecture. Then, in Sec.\ V, the density of proper delay
times is computed using the theory of orthogonal polynomials.

Historically, the Wigner-Smith matrix $Q\propto\partial
S/\partial\varepsilon$ was introduced to study the time evolution
of a wave packet. This application is limited to the average delay
time \cite{FSrev}, essentially because knowledge of the full
time dependence requires also higher derivatives of $S$ than the
first. Still, there exist physical observables that are entirely
determined by $Q$. Some examples that have been discussed in the recent
literature are the capacitance or admittance of a quantum dot \cite{BC},
and the thermopower \cite{langen}.  In addition to the energy derivative
$\partial S/\partial \varepsilon$, one may study the derivative of $S$
with respect to an arbitrary external parameter $X$. Such a parameter
can represent the shape of the cavity, the magnetic field, or a
local impurity potential. Recent examples of parametric derivatives
range from conductance derivatives \cite{brouwer1} and charge pumping
\cite{pump} in quantum dots to spontaneous emission \cite{Missir} and
photo-dissociation \cite{FA} in
optical cavities. The distribution of $\partial S/\partial
X$ can be obtained by methods similar to those used for the derivation of
$\partial S/\partial \varepsilon$. A brief discussion is given in Sec.\
\ref{sec5}. We conclude in Sec.\ \ref{sec7}.

\section{Probability distribution of the scattering matrix and
Wigner's conjecture}
\label{sec2}

In order to describe the energy dependence of the scattering matrix
$S(\varepsilon)$, we make use of the so-called Hamiltonian approach
to chaotic scattering \cite{LW,vwz,GMW}. In this approach, the
$N \times N$ scattering matrix $S(\varepsilon)$ is expressed in terms of an $M
\times M$ random hermitian matrix ${\cal H}$ and a $M \times N$ coupling
matrix $W$,
\begin{equation}
  S(\varepsilon;{\cal H}) = 1 - 2 \pi i W^{\dagger}
    (\varepsilon - {\cal H}  + i \pi W W^{\dagger})^{-1} W.
  \label{eq:SHam}
\end{equation}
The hermitian matrix ${\cal H}$ models the Hamiltonian of the closed cavity,
its dimension $M$ being taken to infinity at the end of the
calculation. The coupling matrix $W$ contains matrix elements between
the scattering states in the leads and the states localized in the
cavity.  We distinguish three symmetry classes, labeled by the
parameter $\beta$: $\beta=1$ ($2$) in the presence (absence) of
time-reversal symmetry and $\beta=4$ in the presence of both
time-reversal symmetry and spin-orbit scattering. The elements of ${\cal H}$
are real (complex, quaternion) numbers for $\beta=1$ ($2$, $4$).
The symmetries of ${\cal H}$ imply the following symmetries for $S$:
$S$ is unitary symmetric ($\beta=1$), unitary ($\beta=2$), or
unitary self-dual ($\beta=4$).

For the problem of chaotic scattering one considers a statistical
ensemble of cavities, which may be obtained by e.g.\ varying the shape
of the cavity.  We ask for the average of some function
$f[S(\varepsilon_1),\ldots,S(\varepsilon_N)]$ over the
ensemble. In the Hamiltonian approach, this
ensemble average is represented by an integration over the
matrix ${\cal H}$, which is taken to be a random hermitian matrix with
probability distribution $P({\cal H})$,
\begin{equation} \label{eq:avg}
  \left\langle f[S(\varepsilon_1),\ldots,S(\varepsilon_N)] \right\rangle
  =
  \int d{\cal H}\, P({\cal H}) f[S(\varepsilon_1;{\cal
H}),\ldots,S(\varepsilon_N;{\cal H}) ].
\end{equation}
Usually ${\cal H}$ is taken from
the Gaussian ensembles from Random Matrix Theory \cite{Mehta},
\begin{equation}
\label{eq2.2}
  P({\cal H}) \propto
    \exp\left(-\sum_{n=1}^M F(E_n)\right),\quad
    F(E)=\frac{\beta\pi^2}{4\,M\,\Delta}\,E^2
\end{equation}
where $E_1,E_2,\ldots,E_M$ are the eigenvalues of ${\cal H}$.
The precise choice of $F(E)$ is not important in the limit
$M \to \infty$ \cite{Brev,HW}.

The ensemble average defined by Eq.\ (\ref{eq:avg}) can also be
formulated in terms of a distribution function of the scattering matrix
$S$:  \begin{equation} \label{eq:avgS}
  \left\langle f[S(\varepsilon_1),\ldots,S(\varepsilon_n)]
  \right\rangle \equiv \int dS_1\ldots dS_n\,
  P(\varepsilon_1,\ldots,\varepsilon_n;S_1,\ldots,S_n)\, f(S_1,\ldots,S_n).
\end{equation}
Here $P(\varepsilon_1,\varepsilon_2,\ldots,\varepsilon_n;S_1,\ldots,S_n)$
is the joint probability distribution of the scattering matrix $S$ at
the energies $\varepsilon_1$, $\varepsilon_2$, \ldots, $\varepsilon_N$.
The measure $dS$
is the unique measure that is invariant under transformations
\begin{equation} \label{eq:circinv}
  S \to V S V'\ ,
\end{equation}
where $V$ and $V'$ are arbitrary unitary matrices ($V' = V^{\rm T}$ for
$\beta=1$ and $V'=V^{\rm R}$ for $\beta=4$ where $\rm T$ denotes the
transpose and $\rm R$ the dual of a matrix).

Statistical averages of the form (\ref{eq:avg}) are completely
characterized by the $N$ eigenvalues of $W^\dagger W$.
For the case of ballistic point contacts (or
``ideal leads''), these eigenvalues are all equal to $M\Delta/\pi^2$
\cite{LW,Brouwer,vwz}. After application of suitable basis transformations
on $S$ and ${\cal H}$ one has therefore $W_{\mu n}=\delta_{\mu
n}(M\Delta)^{1/2}/\pi$ for $\mu=1,\ldots,M$, $n=1,\ldots,N$. In this
case, the distribution $P(\varepsilon,S)$ of $S(\varepsilon)$ at one
single energy is found to be particularly simple \cite{LW,Brouwer},
\begin{equation} \label{eq:constant}
  P(\varepsilon;S) = \mbox{constant}.
\end{equation}
Equation (\ref{eq:constant}) is the starting point of the ``scattering
matrix approach'' to chaotic scattering \cite{Brev}.
The ensemble of scattering matrices that is defined by the probability
distribution (\ref{eq:constant}) and the invariant measure $dS$ is
Dyson's circular ensemble from Random Matrix Theory \cite{Dyson-4}.
Depending on the symmetry class, we distinguish the Circular Orthogonal,
Unitary, and Symplectic Ensembles, (COE, CUE, CSE), for $\beta=1$, $2$,
and $4$ respectively.

The distribution function (\ref{eq:constant})
is invariant under the transformation (\ref{eq:circinv}).  Wigner
\cite{wigner2} conjectured that the unitary invariance extends to the
distribution functional $P[S(\varepsilon)]$, i.e.\ to the whole energy
dependent $S$-matrix ensemble. In other words, Wigner's conjecture is
that (for any $n$) $P(\varepsilon_1,\ldots,\varepsilon_n;S_1,\ldots,S_n)$
is invariant under a simultaneous transformation
\begin{equation}
\label{eq:invn}
  S_j \to V S_j V',\ \ j=1,\ldots,n,
\end{equation}
where $V$ and $V'$ are arbitrary unitary matrices ($V' = V^{\rm T}$,
$V^{\rm R}$ for $\beta=1,4$).

\section{Proof of Wigner's conjecture}
\label{sec3}

Our proof of Wigner's conjecture
consists of two parts. We first present an
alternative random matrix approach for the energy dependence of the
scattering matrix \cite{BB}, and then show that in this approach the
invariance property (\ref{eq:invn}) is manifest.

In the alternative random matrix approach, the role of the $M \times M$
matrix ${\cal H}$ is taken over by an $M \times M$
unitary matrix $U$,
\begin{eqnarray} \label{eq:CUEB}\label{eq:SU}
\label{eq:SStub-4}
  S(\varepsilon) &=& U_{\rm 11} -
  U_{\rm 12} \left(e^{-2 \pi i \varepsilon/M \Delta} +  U_{\rm 22}
  \right)^{-1}
  U_{\rm 21}.
\end{eqnarray}
Here the matrices $U_{\rm 11}$,
$U_{\rm 12}$, $U_{\rm 21}$, and $U_{\rm 22}$ denote four subblocks of $U$,
of size $N \times N$, $N \times (M-N)$, $(M-N) \times N$, and
$(M-N) \times (M-N)$, respectively,
\begin{eqnarray}
  U &=& \left( \begin{array}{ll}
               U_{\rm 11} & U_{\rm 12} \\
               U_{\rm 21} & U_{\rm 22} \end{array} \right)
   \begin{array}{l} \} \ N \\ \} \ M - N\end{array}.
\end{eqnarray}
The energy-dependence of $S$ enters through the phase factor
$\exp({-2 \pi i \varepsilon/M \Delta})$. The matrix $U$ is distributed
according
to the appropriate circular ensemble: $U$ is
distributed according to the circular ensemble, COE (CUE, CSE)
for $\beta=1$ ($2,4$).

We now show that Eq.\ (\ref{eq:SStub-4}) with $M \gg N$
is equivalent to the Hamiltonian approach (\ref{eq:SHam}). Equivalence
of the circular ensemble and the Hamiltonian approach at energy
$\varepsilon=0$ was proven in Ref.\ \onlinecite{Brouwer}. This allows us
to write the $M \times M$ unitary matrix $U$ in terms of an $M \times M$
hermitian matrix ${\cal H}$ and an $M \times M$ coupling matrix $W$,
\begin{equation}
  U = 1 + 2 \pi i W^{\dagger} ({\cal H} - i \pi W
    W^{\dagger})^{-1} W. \label{eq:SWH-4}
\end{equation}
The matrix ${\cal H}$ is distributed according to the
Lorentzian ensemble \cite{Brouwer}, which in the large-$M$ limit is
equivalent to the Gaussian ensemble (\ref{eq2.2}).
[The Lorentzian ensemble has
$F(E)=\frac{1}{2}(\beta M+2-\beta)\,\ln(M^2\,\Delta^2+\pi^2\,E^2)$.]

The coupling matrix $W$
is a square matrix
with elements $W_{\mu n} = \pi^{-1} \delta_{\mu n} (M\,\Delta)^{1/2}$.
We separate $W$ into two
rectangular blocks $W_{\rm 1}$ and $W_{\rm 2}$, of size $M \times N$ and
$M \times (M-N)$, respectively,
\begin{equation}
  W = (W_{\rm 1}, W_{\rm 2}),
\end{equation}
and substitute Eq.\ (\ref{eq:SWH-4}) into
Eq.\ (\ref{eq:SStub-4}). The result is
\begin{equation}
\label{eq:SHam-5}
  S(\varepsilon) = 1 - 2 \pi i W_{\rm 1}^{\dagger}
  (\varepsilon - {\cal H}  - \delta {\cal H}  + i
    \pi W_{\rm 1}^{\vphantom{\dagger}} W_{\rm 1}^{\dagger})^{-1}
  W_{\rm 1}\ ,
\end{equation}
where the $M \times M$ matrix $\delta {\cal H}$ is defined as
\begin{equation}
  \delta {\cal H}  =
     \left( \begin{array}{cc} \varepsilon & 0 \\
            0 & \varepsilon -(M\,\Delta/\pi)\tan(\pi\varepsilon/
                     M\Delta) \end{array} \right)
     \begin{array}{l} \} N \\ \} M-N \end{array}\ .
\end{equation}
Apart from the term $\delta {\cal H}$, Eq.\ (\ref{eq:SHam-5}) is
the same as Eq. (\ref{eq:SHam}) in the Hamiltonian approach.
The extra term $\delta {\cal H}$ is irrelevant in the
limit $M \to \infty$ at fixed $N$, because the diagonal matrix
$\delta {\cal H}$
contains only a finite number of matrix elements that are of order
$\varepsilon$, the others being of order $\varepsilon^2/M \Delta$.
(One easily verifies that such a perturbation does not change
eigenvalues and eigenvectors of ${\cal H}$ in the limit $M\to\infty$.)
This proves equivalence of the alternative random-matrix approach
(\ref{eq:SStub-4}) and the Hamiltonian approach (\ref{eq:SHam}).

It remains to derive the unitary invariance of $P[S(\varepsilon)]$
from Eq. (\ref{eq:SStub-4}). As the matrix $U$ is
chosen from the circular ensemble, its distribution is invariant under
a mapping $U \to U' U U''$, where $U'$ and $U''$ are arbitrary $M
\times M$ unitary matrices ($U'' U'^{*} = 1$ if $\beta=1,4$). We now
choose
\begin{equation} \label{eq:choice}
  U' =   \left( \begin{array}{cc} V & 0 \\ 0 & 1 \end{array} \right)
  \begin{array}{l} \} N \\ \} M-N \end{array},\ \
  U'' =  \left( \begin{array}{cc} V' & 0 \\ 0 & 1 \end{array} \right)
  \begin{array}{l} \} N \\ \} M-N \end{array},
\end{equation}
where $V$ and $V'$ are arbitrary unitary $N \times N$ matrices ($V^{*}
V' = 1$ if $\beta=1,4$).
Invariance of $P(U)$ under the transformation $U \to U' U U''$ with the
choice (\ref{eq:choice}) implies, in view of Eq. (\ref{eq:SStub-4}),
invariance of $P[S(\varepsilon)]$
under the transformation
\begin{equation}
  S(\varepsilon) \to V S(\varepsilon) V',
\end{equation}
as advertized.

\begin{figure}

\hspace{0.15\hsize}
\epsfxsize=0.53\hsize
\epsffile{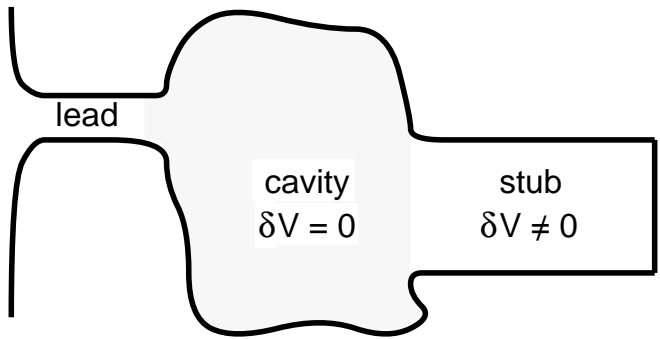}\\

\par

\caption{\label{fig:1-4-1}
\label{fig:2-4-1} Physical motivation of Eq.\ (\protect\ref{eq:SStub-4}),
explained in the text. }
\end{figure}

A central role in our proof of Wigner's conjecture is
played by the random-matrix model (\ref{eq:SStub-4}), which expresses
the energy-dependent $N \times N$ scattering matrix $S(\varepsilon)$
in terms of an energy-independent $M \times M$ random unitary matrix
$U$. This model was first proposed in Ref.\ \onlinecite{BB}, following
a more physical reasoning than the formal derivation given above. It is
insightful to briefly repeat the reasoning of Ref.\ \onlinecite{BB}. We
consider the scattering matrix $S$ at energy
$\varepsilon=0$ and then shift the energy by an amount $\varepsilon$.
Equivalently, we can replace the energy shift $\varepsilon$ by a
uniform decrease $\delta V = -\varepsilon/e$ of the potential $V$ in
the quantum dot.  Since the quantum dot is chaotic, we may localize
$\delta V$ in a closed lead (a stub), see Fig.\ \ref{fig:2-4-1}. The
stub contains $M-N \gg N$ modes to ensure that it faithfully represents
a spatially
homogeneous potential drop $\delta V$. The scattering properties of the
system consisting of the dot, the $N$-mode lead, and the $(M-N)$-mode stub
are described by the $(M-N)$-dimensional reflection matrix
$r_{\rm s}$ of the stub and the $M$-dimensional scattering matrix $U$
of the cavity at energy $\varepsilon=0$, with the stub replaced by an
open lead. The advantage of localizing the potential shift
$\delta V$ inside the stub is that the energy-dependence of $r_{\rm s}$ is
very simple,
\begin{equation} \label{eq:VB}
  r_{\rm s}(\varepsilon) =  -e^{-2 \pi i \varepsilon/M \Delta}.
\end{equation}
The matrix $U$ is energy independent by construction. It is taken from
the circular
ensemble, because scattering from the dot is chaotic.
(The matrix $r_{\rm s}$ is not random.) Expressing the
scattering matrix $S(\varepsilon)$ in terms of $U$ and $r_{\rm
s}(\varepsilon)$ then yields Eq.\ (\ref{eq:SStub-4}).

\section{Distribution of the time-delay matrix}
\label{sec4}

In this section we combine Wigner's conjecture for
the invariance properties of the scattering matrix
ensemble and the Hamiltonian approach to chaotic scattering,
to compute the joint distribution $P(S,Q_\varepsilon)$ of the scattering matrix
$S(\varepsilon)$ and its symmetrized energy derivative
\begin{equation}
\label{eq2.6}
  Q_\varepsilon=-i\hbar S^{-1/2} \frac{\partial S}{\partial \varepsilon}
  S^{-1/2}.
\end{equation}
The matrix $Q_\varepsilon$ is a real symmetric (hermitian, quaternion
self-dual) matrix for $\beta=1$ ($2$, $4$).
The joint distribution $P(S,Q_\varepsilon)$ of $S$ and $Q_\varepsilon$ is
defined through a relation similar to Eqs.\ (\ref{eq:avg}) and (\ref{eq:avgS}),
\begin{eqnarray}
  \langle f[S(\varepsilon),Q_\varepsilon(\varepsilon)] \rangle &=&
  \int d{\cal H}\, P({\cal H}) f[S(\varepsilon,{\cal H}),
  Q_\varepsilon(\varepsilon,{\cal H})] \nonumber \\
  & \equiv & \int dS\, dQ_\varepsilon\, f(S,Q_\varepsilon)
  P(S,Q_\varepsilon), \label{eq:PSQdef}
\end{eqnarray}
where $f(S,Q_\varepsilon)$ is an arbitrary function of $S$ and $Q_\varepsilon$.

Application of Wigner's conjecture is the key to the calculation
of $P(S,Q_\varepsilon)$. To see this, we first consider the
distribution function $P(\varepsilon_2-\varepsilon_1;S_1,S_2)$ of
the scattering matrix $S(\varepsilon)$ at the energies $\varepsilon_1$
and $\varepsilon_2$, for
which the invariance property (\ref{eq:invn}) implies
\begin{eqnarray}
\label{eq2.5}
  P(\varepsilon_2-\varepsilon_1;S_1,S_2) &=&
  P(\varepsilon_2-\varepsilon_1;\,V S_1 V', V S_2 V') \nonumber \\ &=&
  P(\varepsilon_2-\varepsilon_1;-1,-S_1^{-1/2}\,S_2\,S_1^{-1/2}).
\end{eqnarray}
For the second equality we have chosen $V=V'=i\,S_1^{-1/2}$. In the
limit $\varepsilon_2\to \varepsilon_1=\varepsilon$, one finds from
Eq.\ (\ref{eq2.5}) that the joint distribution
$P(S,Q_\varepsilon)$ of the scattering matrix $S(\varepsilon)$ and the
symmetrized time-delay matrix $Q_\varepsilon$ does not depend on the
scattering matrix $S$,
\begin{equation}
\label{eq2.8}
  P(S,Q_\varepsilon)=P(-1,Q_\varepsilon).
\end{equation}
Hence, to find $P(S,Q_\varepsilon)$ it is sufficient to calculate the integral
(\ref{eq:PSQdef}) for a function $f(S,Q_\varepsilon)$ of the form
\begin{equation} \label{eq:special}
  f(S,Q_\varepsilon) =
\delta(S,\,-1) f(Q_\varepsilon),
\end{equation}
where $f(Q_\varepsilon)$ is an arbitrary function of $Q_\varepsilon$ and
$\delta(S,\,-1)$ is
the delta function at $S=-1$ on the manifold of unitary
symmetric (unitary, unitary self-dual) matrices for $\beta=1$ ($2$, $4$).

We now turn to the evaluation of the integral (\ref{eq:PSQdef}), with
$f$ of the form (\ref{eq:special}), using
the Hamiltonian approach. We note that (\ref{eq:SHam}) can be rewritten
as
\begin{equation}
\label{eq4.1} \label{eq4.2}
  S = -1+\frac{2}{1+i\,K}, \ \
  K = \pi\,W^\dagger\frac{1}{\varepsilon-{\cal H}}\,W\ .
\end{equation}
The energy derivative of $S$ is given by
\begin{equation}
\label{eq4.3}
\frac{\partial S}{\partial \varepsilon}=2\pi i\,\frac{1}{1+i\,K}
\,W^\dagger\frac{1}{(\varepsilon-{\cal H})^2}\,W\,\frac{1}{1+i\,K}\ .
\end{equation}
We decompose the hermitian matrix ${\cal H}$ in Eq.\ (\ref{eq:PSQdef}) into
its eigenvalues and eigenvectors: ${\cal H}=\psi E\psi^\dagger$, where $E$
is the diagonal matrix containing the eigenvalues $E_\rho$ of ${\cal H}$ as
entries and $\psi$ is the orthogonal (unitary, symplectic) matrix of
the eigenvectors of ${\cal H}$ for $\beta=1$ ($2$, $4$),
\begin{eqnarray}
\nonumber
\langle f(S,Q_\varepsilon)\rangle &=& \int d\psi \int dE_1\ldots dE_M\,
J_M(E) P(E)\, f[S(\varepsilon;\psi E\psi^\dagger),Q_\varepsilon
(\varepsilon;\psi E\psi^\dagger)], \\ J_M(E) &=&
\prod_{\rho < \sigma}^{M} |E_\rho-E_{\sigma}|^\beta
\label{eq4.4}.
\end{eqnarray}
Here $d\psi$ is the invariant measure on the orthogonal (unitary, symplectic)
group and $J_M(E)$ is the jacobian for the variable transformation
${\cal H} \to \psi,E$ \cite{Mehta}.

We now make the special choice
(\ref{eq:special}) for $f(S,Q_\varepsilon)$. In terms of the matrices $\psi$
and
$E$, the matrix $K$ from Eq.\ (\ref{eq4.2}) reads
\begin{equation}
\label{eq4.5}
K_{mn}=\frac{M\Delta}{\pi}
\sum_{\rho=1}^M\,
\frac{\psi_{m\rho}\,\psi_{n\rho}^*\,}{\varepsilon-E_\rho}\ .
\end{equation}
Inspecting (\ref{eq4.1}), we see that the limit $S\to-1$ corresponds
to the case when all $N$ eigenvalues of $K$ diverge or,
equivalently, when at least $N$ out of $M$ eigenvalues $E_\rho$ of ${\cal H}$
tend to the energy $\varepsilon$. Let us label these
eigenvalues $E_1,\ldots,E_N$, i.e.
$|E_\rho-\varepsilon|\to 0$ in the limit $S\to-1$ for $\rho=1,\ldots,N$.
In this limit, the sum in (\ref{eq4.5}) can be restricted to the first
$N$ terms and both $S$ and $\partial S/\partial \varepsilon$
do not depend on the
eigenvalues $E_{\rho}$ and on the matrix elements $\psi_{m \rho}$ with
$m > N$ or $\rho>N$.
Therefore we can perform the integration with respect to the latter,
resulting in
\begin{eqnarray}
\label{eq:SN}
  \langle \delta(S,-1)f(Q_\varepsilon)\rangle &=&
  \int dQ_\varepsilon f(Q_\varepsilon) P(-1,Q_\varepsilon) \nonumber \\
  &=&\int d\Psi P(\Psi) \int dE_1\ldots dE_N\, P(E_1,\ldots,E_N)\,
\delta(S,-1) f(Q_\varepsilon),
\end{eqnarray}
where $\Psi$ is the $N \times N$ matrix containing the rescaled
eigenvector elements $\psi_{m\rho} M^{1/2}$ for $1 \le m,\rho \le N$,
the integration measure $d\Psi = \prod_{m,\rho=1}^{N} d\Psi_{m\rho}$,
$P(\Psi)$ is the
distribution of $\Psi$ after integration over the remaining matrix
elements of $\psi$, and $P(E_1,\ldots,E_N)$ is the distribution of the
eigenvalues $E_1$,\ldots,$E_N$ after integration over the remaining
eigenvalues $E_{\rho}$, $\rho>N$.
Near $S=-1$ we have from Eqs.\ (\ref{eq4.2}), (\ref{eq4.3}), and
(\ref{eq4.5}),
\begin{eqnarray}
  S &=& -1 + (2 \pi i/\Delta) \Psi^{\dagger-1}
  \mbox{diag}\,(E_1-\varepsilon,\ldots,E_N-\varepsilon)
  \Psi^{-1},\\ Q_\varepsilon &=& \tau_H \Psi^{\dagger-1} \Psi^{-1},
\end{eqnarray}
where $\tau_H = 2 \pi \hbar/\Delta$ is the Heisenberg time.
In the limit $M \gg N$, the matrix elements of $\Psi$ are Gaussian
distributed with zero mean and unit variance
\cite{Mehta},
\begin{equation}
  P(\Psi) \propto \exp[-(\beta/2) \mbox{tr}\, \Psi \Psi^{\dagger}].
\end{equation}
Because of the delta function $\delta(S,-1)$, it is sufficient to know
$P(E_1,\ldots,E_N)$ for $|E_{\rho} - \varepsilon| \to 0$, $\rho=1,
\ldots,N$. In this case, $P(E_1,\ldots,E_N)$ is entirely determined by
the Jacobian $J_N(E_1,\ldots,E_N)$, up to a numerical factor which
results from the integration over the remaining $M-N$
eigenvalues of ${\cal H}$,
\begin{equation}
  P(E_1,\ldots,E_N) \propto \prod_{\rho<\sigma}^{N} |E_{\rho} -
  E_{\sigma}|^{\beta}\ .
\end{equation}

We have now succeeded in transforming the integral (\ref{eq:PSQdef})
over an $M\times M$ matrix ${\cal H}$ to the
integral (\ref{eq:SN}), which is formulated in terms of $N \times N$
matrices only.  At this point, it is convenient to rewrite the
integration over the $N$ eigenvalues $E_1,\ldots,E_N$ as a matrix
integration to avoid the singularity of $P(E_1,\ldots,E_N)$
at $E_1 = \ldots = E_N = \varepsilon$.
Hereto we exploit that the distribution of $\Psi$ is invariant with
respect to a transformation of the type $\Psi\to O\Psi$,
$O$ being an orthogonal (unitary, symplectic) matrix for $\beta=1$
($2$, $4$). We apply this transformation for arbitrary $O$
and average over $O$ with respect to the invariant measure
$dO$. Then, we substitute the hermitian
$N\times N$-matrix
$H\equiv O^\dagger \mbox{diag}\,(E_1,\ldots,E_N) O - \varepsilon$
and the corresponding
Jacobian exactly cancels the singular level-repulsion factor of
$P(E_1,\ldots,E_N)$,
\begin{eqnarray}
  \langle \delta(S,-1) f(Q_\varepsilon)\rangle & \propto &
    \int d\Psi P(\Psi) \int dH\, \delta(S,-1) f(Q_\varepsilon)
  \nonumber \\ & \propto &
    \int d\Psi P(\Psi) \int dH\,
    \delta(\Psi^{\dagger-1}\,H\,\Psi^{-1})
    f[\tau_H \Psi^{\dagger-1}\Psi^{-1}].
  \label{eq4.11}
\end{eqnarray}
(The proportionality sign indicates that we have omitted a normalization
constant.)
Three more variable transformations are needed to calculate the
integral (\ref{eq4.11}).
First, we replace the integration variable $H$ by $H' =
\Psi^{\dagger-1}\,H\,\Psi^{-1}$. The Jacobian
$\det(\Psi\,\Psi^\dagger)^{(N\beta + 2 - \beta)/2}$
of this transformation can be derived using the singular value
decomposition of $\Psi$ \cite{FM}.
The integral (\ref{eq4.11}) becomes
\begin{eqnarray}
  \langle \delta(S,-1) f(Q_\varepsilon)\rangle & \propto &
    \int d\Psi P(\Psi) \int dH'\,
    \delta(H') \det(\Psi\,\Psi^\dagger)^{(N\beta + 2 - \beta)/2}
    f[\tau_H \Psi^{\dagger-1}\Psi^{-1}] \nonumber \\ &=&
  \int d\Psi e^{-(\beta/2) \mbox{tr}\, \Psi \Psi^{\dagger}}
    \det(\Psi\,\Psi^\dagger)^{(N\beta + 2 - \beta)/2}
    f[\tau_H \Psi^{\dagger-1}\Psi^{-1}].
\end{eqnarray}
Next, we note that the integrand depends on the hermitian matrix
$\Gamma\equiv \Psi\Psi^\dagger/\tau_H$ only, and replace the
integration variable $\Psi$ by $\Gamma = \Psi\Psi^{\dagger}/\tau_H$.
The Jacobian of this transformation \cite{SlevinNagao,BrezinHikamiZee}
provides an extra factor
$\propto\det(\Gamma)^{\beta/2-1} \theta(\Gamma)$, where
$\theta(\Gamma)=1$ if all eigenvalues of $\Gamma$ are positive and
$\theta(\Gamma) = 0$ elsewise.
Finally, we replace $\Gamma$ by its inverse $Q = \Gamma^{-1}$. Since
this variable transformation has a Jacobian
$\det (Q)^{\beta N + 2 -\beta}$, we arrive at
\begin{eqnarray}
  \langle \delta(S,-1) f(Q_\varepsilon)\rangle & \propto &
    \int dQ\, \theta(Q) \det(Q)^{-3 \beta N/2 - 2 + \beta} e^{-\beta \tau_H
    {\rm tr}\, Q^{-1}/2} f(Q).
\end{eqnarray}
Using Eqs.\ (\ref{eq2.8}) and (\ref{eq:SN}) we thus find the joint
probability distribution $P(S,Q_\varepsilon)$ of the scattering matrix $S$
and the symmetrized time delay matrix $Q_\varepsilon$,
\begin{eqnarray}
  P(S,Q_\varepsilon) = P(-1,Q_\varepsilon) = \theta(Q_\varepsilon)
\det(Q_\varepsilon)^{-3 \beta N/2 - 2 + \beta} e^{-\beta \tau_H
    {\rm tr}\, Q_\varepsilon^{-1}/2}.
\end{eqnarray}

The corresponding distribution of the eigenvalues
$\tau_1,\ldots,\tau_N$ of $Q_\varepsilon$, the proper delay times then reads
\begin{equation}
\label{eq4.17}
P(\tau_1,\ldots,\tau_N)\propto
\prod_{n<m}|\tau_n-\tau_m|^\beta\ \prod_n
\theta(\tau_n)\,\tau_n^{-3\beta N/2-2+\beta}
\ e^{-\beta \tau_H/(\tau_n\,2)}.
\end{equation}
Alternatively, the distribution of the rates $\gamma_j = 1/\tau_j$
is given by
\begin{equation}
\label{eq4.16}
P(\gamma_1,\ldots,\gamma_N)\propto
\prod_{n<m}|\gamma_n-\gamma_m|^\beta\ \prod_n
\theta(\gamma_n)\,\gamma_n^{\beta N/2}\ e^{-\beta \tau_H\gamma_n/2}\ .
\end{equation}
This distribution is known in random-matrix theory
as the generalized Laguerre ensemble \cite{Mehta}.
The eigenvectors of $Q_\varepsilon$
are uniformly distributed according to the invariant measure of the
orthogonal (unitary, symplectic) group.

\section{Density of proper delay times}
\label{sec6}

The distributions (\ref{eq4.17}) and
(\ref{eq4.16}) have a form which allows to apply
the standard technique of orthogonal polynomials \cite{Mehta}
to determine the density or the correlation functions
of the eigenvalues of $Q_\varepsilon$ (or $\Gamma = Q_\varepsilon^{-1}$).
The $N$-dependent exponent is
somewhat unusual and implies that the set of orthogonal polynomials
is different for each value of $N$. Such ensembles have
been studied in mathematical physics \cite{Edelman,Baker}.

We restrict ourselves to the simplest case $\beta=2$ of broken
time reversal symmetry. For a simplified notation,
we use the dimensionless escape rates $x_n\equiv \gamma_n\,\tau_H =
\tau_H/\tau_n$ which are distributed according to
\begin{equation}
\label{eq6.1}
p(x_1,\ldots,x_N)\propto\prod_{n<m}(x_n-x_m)^2\,\prod_n w_N(x_n)\quad,
\quad w_N(x)=x^N\,e^{-x}\ .
\end{equation}
The generalized Laguerre polynomials $L_{n}^{N}(x)$ are orthogonal with
respect to the weight function $w_{N}(x)$. The method of orthogonal
polynomials relates the correlation functions of the $x_{n}$'s to these
polynomials. Here we only consider the density
\begin{equation}
\rho(x)=\int_{0}^{\infty} dx_{1}\ldots dx_{N}\ p(x_{1},\ldots,x_{N})
\sum_{n=1}^{N}\delta(x-x_{n}),
\end{equation}
which is given by the series
\begin{equation}
\label{eq6.6}
  \rho(x) =\sum_{n=0}^{N-1}
  {L_n^{(N)}(x)^2\,w_N(x)\,n!\over (N+n)!} .
\end{equation}
Using the recurrence relation \cite{abramowitz}
\begin{equation}
\label{eq6.7}
(n+1)\,L_{n+1}^{(N)}(x)=(2n+N+1-x)\,L_n^{(N)}(x)-(n+N)\,L_{n-1}^{(N)}(x)\ ,
\end{equation}
with $L_0^{N}(x)=1$ and $L_1^{(N)}(x)=N+1-x$, it is possible
to evaluate Eq. (\ref{eq6.6}) efficiently for small $N$.
For large $N$, one can use an asymptotic expansion of the Laguerre
polynomials to find the closed expression
\begin{equation}
\rho(x) =
  \frac{1}{2\pi x}\sqrt{-N^2+6Nx-x^2}\ .
\label{eq6.15}
\end{equation}
The corresponding density for the proper delay times
$\tau=\tau_H/x$ thus reads
\begin{equation}
\label{eq6.16}
\rho(\tau)=\frac{N}{2\pi\tau^2}
\sqrt{(\tau_+-\tau)(\tau-\tau_-)}\quad,\quad
\tau_\pm=\frac{\tau_H}{N}(3\pm \sqrt{8})\ .
\end{equation}
The limit $N\rightarrow\infty$ is rapidly approached for $N\gtrsim 3$, see
Fig.\ 2.

\begin{figure}

\hspace{0.15\hsize}
\epsfxsize=0.53\hsize
\epsffile{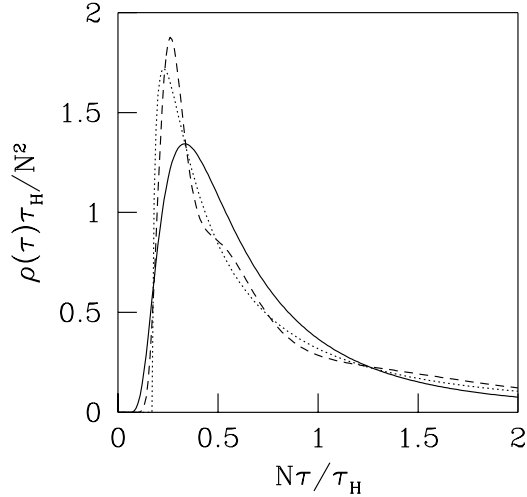}

\caption{\label{fig6.1}
Density of proper delay times.  Shown are the densities
for $N=1$ (full line), $N=3$ (dashed line), both computed from Eq.\
(\ref{eq6.6}), and the limit $N\to\infty$ (dotted line) given by Eq.\
(\ref{eq6.16}).}
\end{figure}

\begin{figure}

\hspace{0.15\hsize}
\epsfxsize=0.53\hsize
\epsffile{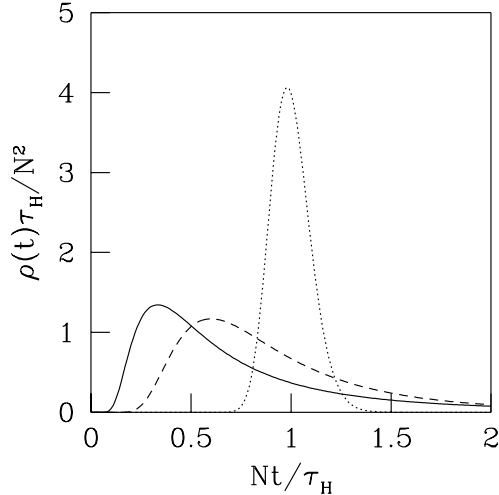}

\caption{\label{fig6.2} Density of partial delay times, from Ref.\
\protect\cite{FS1}. Shown are the densities for $N=1$
(full line), $N=3$ (dashed line) and $N=100$ (dotted line).
The limit $N\gg 1$ corresponds to a gaussian with mean $1$ and width
$1/\protect\sqrt{N}$.}
\end{figure}

It is instructive to compare the density of the proper
delay times $\tau_{n}$ with the density of partial delay times studied
by Fyodorov, Sommers, and Savin \cite{FSrev,FS1}.  The partial delay
times $t_{n}$ are defined in terms of the eigenvalues $e^{i \phi_n}$
of the scattering matrix $S$,
\begin{equation}
\label{eq6.17}
t_n=\hbar\frac{\partial \phi_n}{\partial \varepsilon}.
\end{equation}
Compared to the definition of proper delay times as
eigenvalues of the Wigner-Smith matrix, we see that for the partial
delay times the order of the operations of energy derivative and
diagonalization is reversed. This order is irrelevant for the sum
$\sum_{n}\tau_{n}=\sum_{n}t_{n}$, corresponding to the first moment
of the density, but higher moments are different (unless $N=1$). The
density of partial delay times was obtained in
Ref.\ \cite{FSrev,FS1} by the supersymmetric approach,
\begin{equation}
\label{eq6.18}
\rho(t)=\frac{1}{(N-1)!\,\tau_H}\,\left(\frac{\tau_H}{t}\right)^{N+2}
\,e^{-\tau_H/t}\ .
\end{equation}
Notice that this density corresponds to the contribution from the
first Laguerre polynomial in the summation (\ref{eq6.6}). For $N\ge 2$,
there are also contributions from higher order Laguerre polynomials and
therefore the two densities do not coincide.
In Fig.\ \ref{fig6.2}, the density of the partial delay times is shown for
$N=1,\,3,\,100$.
For $N\gg 1$, Eq.\ (\ref{eq6.18}) becomes a gaussian of mean $1$
and a width $1/\sqrt{N}$. The qualitative difference between $\rho(\tau)$
and $\rho(t)$ is due to the absence of level repulsion for the partial
delay times. Unlike the proper delay times, we do not know of a physical
quantity that is determined by the partial delay times.

\section{General parametric derivatives}
\label{sec5}

So far we have restricted our discussion to the energy derivative of
the scattering matrix. It is also of interest to know the derivatives
with respect to external parameters. Examples of such parameters are a
magnetic field, the shape of the cavity, or a local potential.  In this
section we present and prove an extension of Wigner's conjecture to the
ensemble of scattering matrices $S(\varepsilon,X)$ that depend both on
energy and on an external parameter $X$, and then use it to compute the
distribution of the symmetrized derivative
\begin{equation}
  Q_X = -i S^{-1/2} {\partial S \over \partial X} S^{-1/2}.
\end{equation}

In the Hamiltonian approach, the parameter dependence
of the scattering matrix is modeled through Eq.\ (\ref{eq:SHam}) with
a parameter dependent random hermitian matrix ${\cal H}(X)$,
\begin{eqnarray}
  S(\varepsilon,X;{\cal H}) &=& 1 - 2 \pi i W^{\dagger}
    [\varepsilon - {\cal H}(X)  + i \pi W W^{\dagger}]^{-1} W, \nonumber \\
  \label{eq:SHamX}
 {\cal H}(X) &=& {\cal H}  + X {\cal H}'. \label{eq:HX}
\end{eqnarray}
The matrix ${\cal H}'$ is a random hermitian, Gaussian distributed, symmetric
(antisymmetric) matrix if $X$ denotes a shape change (magnetic field),
or ${\cal H}'_{ij} = \delta_{ir} \delta_{jr}$ if $X$ denotes the local
potential at site $r$. (In random-matrix theory, the site $r$ is
represented by an index $1 \le r \le M$, see
e.g.\ Ref.\ \onlinecite{Missir}.) As we argue below, the invariance
property (\ref{eq:invn}), which was conjectured by Wigner for the
ensemble of energy-dependent scattering matrices, is also valid for the
more general ensemble of parameter- and energy-dependent scattering
matrices $S(\varepsilon,X)$. That is to say, the distribution functional
$P[S(\varepsilon,X)]$ of the ensemble of matrix valued functions
$S(\varepsilon,X)$ is invariant under transformations
\begin{equation} \label{eq:invariant32}
  S(\varepsilon,X) \to V S(\varepsilon,X) V',
\end{equation}
where $V$ and $V'$ are arbitrary unitary matrices ($V^{*} V' = 1$ if
$\beta=1,4$).
The proof of this invariance property goes along similar lines as in
Sec.\ \ref{sec3}:
First, one shows that the Hamiltonian approach (\ref{eq:SHamX}) for the
ensemble of parameter dependent scattering matrices is equivalent to a
formulation \onlinecite{BrouwerBeenakker1996}
in terms of unitary matrices of a form similar to
Eq.\ (\ref{eq:SStub-4}),
\begin{eqnarray}
  S(\varepsilon,X) &=& U_{\rm 11} +
  U_{\rm 12} \left[e^{-2 \pi i (\varepsilon - X {\cal H}')/M \Delta} -
  U_{\rm 22} \right]^{-1}
  U_{\rm 21}. \label{eq:SStub-5}
\end{eqnarray}
Second, one verifies that in the
formulation (\ref{eq:SStub-5}) the invariance property
(\ref{eq:invariant32}) is manifest, completing the proof.

Using the invariance property (\ref{eq:invariant32}), we can now compute the
distribution of $Q_X$. For simplicity, we restrict
ourselves to the case that the matrices ${\cal H}$ and ${\cal H}'$
have the same,
Gaussian, distribution.  As in the case of the energy derivative of the
scattering matrix, see Sec.\ \ref{sec4}, it is sufficient to consider
the special point $S=-1$. {}From Eqs.\ (\ref{eq4.2}) and
(\ref{eq:HX}) we find that at $S = - 1$
\begin{mathletters}
\begin{eqnarray}
  Q_X &=& \Psi^{\dagger-1} H' \Psi^{-1},
  \ \ P(H') \propto e^{-\beta \mbox{tr}\,H'^2/16},\\
  H'_{mn} &=& -(\tau_H/\hbar) M^{-1/2} \sum_{\mu, \nu} \psi_{m \mu}^{*}
  {\cal H}'_{\mu \nu} \psi_{n \nu}^{\vphantom{*}}.
\end{eqnarray}
\end{mathletters}%
Repeating the steps outlined
in the previous section, we find that the distribution of $Q_X$
is Gaussian with a width set by $Q_\varepsilon$,
\begin{eqnarray}
  && P(S,Q_\varepsilon,Q_X) \propto (\det Q_\varepsilon)^
  {-2\beta N-3+3\beta/2}\,
  \exp\left[{-{\beta \over 2}\,\mbox{tr}\,
    \left(\tau_H Q_\varepsilon^{-1} +
    {1 \over 8}(\tau_H Q_\varepsilon^{-1} Q_X^{\vphantom{1}})^2\right)}\right].
    \label{eq:PSQQ}
\end{eqnarray}
The reason why the time-delay matrix $Q_\varepsilon$ sets the scale for the
matrix $Q_X$ characterizing the response to the external paramater
$X$ can be understood in a picture of classical trajectories
\cite{jalabert}: For long delay times, the scattering properties are
more senstive to a perturbation of the system than for short delay
times. Such an explanation in
terms of classical trajectories is valid in the semiclassical limit $N
\to \infty$ only.
Our exact result (\ref{eq:PSQQ}) makes the relation between time delay
and parameter response precise in the fully quantum mechanical regime
of a small number of channels $N$.

The invariance property (\ref{eq:invariant32}) and the Gaussian
distribution (\ref{eq:PSQQ}) also hold if the scattering matrix depends
on more than one external parameter. If the scattering matrix depends
on parameters $X_1$,\ldots, $X_n$, the matrices $Q_{X_j}$
($j=1,\ldots,n$) all have the same Gaussian distribution, with a
width set by the symmetrized time-delay matrix $Q_\varepsilon$,
\begin{eqnarray}
  P(S,Q_\varepsilon,Q_{X_1},\ldots,Q_{X_n}) &\propto&
  (\det Q_\varepsilon)^{-N/2 - (\beta N+2-\beta)(n+2)/2}\,
  \nonumber \\ && \mbox{} \times
  \exp\left[{-{\beta \over 2}\,\mbox{tr}\,
    \left(\tau_H Q_\varepsilon^{-1} +
    {1 \over 8}\sum_{j=1}^{n}(\tau_H Q_\varepsilon^{-1}
    Q_{X_j}^{\vphantom{1}})^2\right)}\right].
\end{eqnarray}

\section{Conclusion}
\label{sec7}

In summary, we have
presented a detailed derivation of the joint distribution
$P(S,Q_\varepsilon)$ of the
scattering matrix $S$ and the symmetrized time delay
matrix
$Q_\varepsilon = -i \hbar S^{-1/2} (\partial S/\partial \varepsilon) S^{-1/2}$
of a chaotic cavity with ideal leads.
Our result, which was first reported in Ref.\ \onlinecite{BFB}, reads
\begin{eqnarray}
\label{eq7.1}
  P(S,Q_\varepsilon) \propto \theta(Q_\varepsilon)
  \det(Q_\varepsilon)^{-3 \beta N/2 - 2 + \beta}
  \ e^{-\beta \tau_H
    {\rm tr}\, Q_\varepsilon^{-1}/2},
\end{eqnarray}
where $\tau_H = 2 \pi \hbar/\Delta$ is the Heisenberg time and $\Delta$
is the mean level spacing in the cavity. The distribution $P(S,Q_\varepsilon)$
depends on the eigenvalues
$\tau_1,\ldots,\tau_N$ of $Q_\varepsilon$ only; it does not depend
on $S$, nor on the eigenvectors of $Q_\varepsilon$.
Our derivation was based on an old conjecture by Wigner \cite{wigner2},
which we have proven in this paper, that the distribution functional
$P[S(\varepsilon)]$ of the ensemble of energy-dependent scattering
matrices is invariant under unitary transformations. We generalized
Wigner's conjecture to the dependence of $S$ on an external parameter $X$
and derived the distribution of parametric derivatives.

Throughout this paper we assumed ideal coupling of the chaotic cavity to
the electron reservoirs. This requires ballistic point contacts.
For the case of non-ideal coupling,
i.e.\ if the point contacts contain a tunnel barrier, the distribution of the
scattering matrix $S$
is given by the more general Poisson kernel instead of the
circular ensemble \cite{Brouwer,MPS}.
For this situation, application of Wigner's conjecture is not
correct and our method can not be used to determine $P(S,Q_\varepsilon)$.
A possibility to overcome this difficulty is to separate the
direct and the fully ergodic scattering processes,
by expressing the scattering matrix for non-ideal coupling as a
composition of an $N \times N$ scattering matrix $S_{\rm erg}$
of a fully ergodic and chaotic scatterer and a scattering matrix
$S_{\rm dir}$ of dimension $2N\times 2N$ describing the direct
scattering processes \cite{Brouwer,FM,GM},
\begin{equation}
  S = r + t' (1 - S_{\rm erg} r')^{-1}
      S_{\rm erg} t,\ \
\end{equation}
where the matrices $r$, $r'$, $t$, and
$t'$ are $N \times N$ submatrices of $S_{\rm dir}$,
\begin{equation}
  S_{\rm dir} = \left( \begin{array}{ll} r & t' \\ t & r' \end{array} \right).
\end{equation}
The energy dependence of $S_{\rm dir}$ can
be neglected on energy scales corresponding to the inverse dwell time
inside the cavity. Thus one obtains a relation between the statistical
distributions of the time delay matrices for the cases of
non-ideal and ideal coupling. What remains is the cumbersome mathematical
problem to perform the corresponding transformation of the different
matrix variables. To our knowledge, no closed-form formula for the joint
distribution of $S$ and $Q_\varepsilon$ is known, except for the case $N=1$
\cite{GM}.

The supersymmetric approach used in Refs.
\cite{FSrev,Leh,FS1,vwz}
is typically more general and systematic concerning the issue
of non-ideal coupling, since there the particular symmetries of the
scattering matrix ensemble in the ideal coupling case
are not directly exploited. On the other hand, this technique
is usually restricted to the computation of certain Green function averages
only, and does not give access to the full distibution of the time-delay
matrix. We believe that, in principle, in the supersymmetric approach
it should be possible to
determine averages of the form $\langle \mbox{tr}\, (Q^n) \rangle$ for
arbitrary $n$ (and hence the density of delay times) and for
arbitrary coupling, without having to increase the dimension of the
supermatrices involved. However, in order to get access to the full
distribution of the matrix $Q_\varepsilon$, or at least to moments of the
type $\langle (\mbox{tr}\, Q)^n\rangle$, one has to increase the dimension
of the supermatrices accordingly, which makes it apparently impossible
to go beyond the case $n=2$ \cite{FSrev,Leh,FS1}.
It is amusing that the (now proven)
half-a-century old conjecture of Wigner provides a route to the simple
result (\ref{eq7.1}) that seems unreachable by later supersymmetric
techniques.

\acknowledgments

This research was supported by the ``Ne\-der\-land\-se or\-ga\-ni\-sa\-tie
voor We\-ten\-schap\-pe\-lijk On\-der\-zoek'' (NWO) and by the
``Stich\-ting voor Fun\-da\-men\-teel On\-der\-zoek der Ma\-te\-rie''
(FOM). PWB also acknowledges the support of the NSF
under grants no. DMR 94-16910, DMR 96-30064, and DMR 97-14725.

\end{document}